\documentstyle[aps,pre,preprint]{revtex}
\begin{document}
\draft
\title{Velocity distribution of fluidized granular gases
in presence of gravity}
\author{J. Javier Brey and  M.J. Ruiz--Montero}
\address{F\'{\i}sica Te\'{o}rica, Universidad de Sevilla, Apdo.\ de
 Correos 1065, E-41080 Sevilla, Spain}
\date{today}

\maketitle

\begin{abstract}

The velocity distribution of a fluidized dilute granular gas 
in the direction perpendicular to the gravitational field is investigated
by means of Molecular Dynamics simulations. The results indicate that 
the velocity distribution can be exactly described neither by
a Gaussian nor by a stretched exponential law. Moreover, it does not
exhibit any kind of scaling. In fact, the actual
shape of the distribution depends on the number of monolayers at
rest, on the restitution coefficient and on the height at what
it is measured. The role played by the number of particle-particle
collisions as compared with the number of particle-wall collisions is
discussed.

\end{abstract}


\section{Introduction}
\label{s1}

Steady states of  granular fluids require that energy be continuously 
supplied to the system, in order to compensate the energy loss in collisions.
In many experimental situations, this is done by means of a vibrating wall.
If the vibration is strong enough, the steady 
state will correspond to a completely  fluidized situation. 
The velocity distribution function (VDF) of a granular fluid
in such a steady state has attracted a lot of attention in recent 
years. It has been studied theoretically, numerically, and by means of 
experiments, using very different drivings of the system. 
It seems clear at this stage that the results are strongly dependent on the 
characteristics of the driving mechanism, 
and also on the granular fluid itself. In this paper, we 
will be dealing with the  VDF of a fluidized low density granular 
system in the presence of gravity, 
and in a situation when the number of wall-particle collisions 
is much smaller than the number of particle-particle collisions. 
In these conditions, one can expect that the VDF will be dominated by 
the intrinsic dynamics of the granular gas, and not by the peculiarities 
of the energy injection.

	One of the first experimental studies of the VDF of a granular
fluid in the  situation described above  was carried out  by Warr {\em et al}
 \cite{WHyJ95}. They considered a vertical two-dimensional system of 
steel balls under vertical
sinusoidal vibration, and found that the width of the VDF for the 
vertical ($v_{y}$) and horizontal ($v_{x}$) components of the velocity   
decreased as height increased,  showing that the granular temperature was
a  decreasing function of height. Besides, the distribution for $v_{x}$ seemed
to be quite well fitted by a gaussian distribution. 
A similar result was found by Helal {\em et al} \cite{HByH97}, 
who performed Molecular Dynamics simulations of a two dimensional system 
of hard disks, concluding again that the distribution function for $v_{x}$ 
was gaussian, at least in the  thermal velocity region. A few years later,
Kudrolli and Henry \cite{KyH00} studied experimentally 
a system very similar to the one in
Ref. \cite{WHyJ95}, but now the plane of the system formed a certain angle 
$\theta$ with the horizontal direction.
By varying $\theta$, the external (gravitational) field was modified, 
allowing to study the effect of varying the particle-particle and 
particle-wall number of collisions ratio. As the steady state
exhibited gradients in the vertical direction, they studied the VDF in
a region chosen so that density varied weakly inside it. They found that the 
width of the distribution for $v_{x}$ increased with $\theta$, and that
for the larger values of $\theta$ studied, the distribution was approximately
gaussian. The authors claimed that this happened because increasing $\theta$
increases the frequency of particle-wall collisions, resulting
in broader distributions.

Rouyer and Menon \cite{RyM00} have reported the results of 
 another experimental study of a two dimensional system of steel balls 
in the presence of gravity.   
To avoid the effects of the inhomogeneities in the VDF, they placed the system
between two identical vibrating walls, one located  at the top and the other
at the bottom of the system. For fast 
enough shaking, a symmetrical around the center of mass state was generated,
with small gradients in the central region. The VDF was measured 
in that central region, and it was found that  the distribution for 
the horizontal component of the velocity, when scaled with its width 
$\sigma_{x}$,  showed an universal behavior. Moreover, it
could be fitted in the whole velocity range by means of  a stretched 
exponential type of distribution, 
\begin{equation}
R(c_{x}) =C^{-1} exp\{-A |c_{x}|^{1.5}\} \,  , 
\label{eq1:1}
\end{equation}
where $c_{x}=v_{x}/\sigma_{x}$, while $A$ and $C$ are constants that can be 
determined from the normalization condition and the fact that the second
moment must be unity, resulting $A\simeq 0.797$, $C\simeq 2.101$. It must be
noticed that, although the same type of behavior had been previously
justified theoretically \cite{VNyE98}, it appeared in the context of
uniformly heated systems, and was only displayed by the high energy
tails of the velocity distribution. In this sense, it is remarkable that,
for all the values of the parameters considered by Rouyer and Menon, 
and in the whole velocity range analyzed,
the gaussian distribution provided always a worse fit to the experimental data
than $R(c_{x})$.  On the other hand, previous studies had found a 
gaussian behavior, at least for some range of the parameters 
\cite{WHyJ95,HByH97,KyH00}, as mentioned above. 

 One of the most important results reported in  \cite{RyM00} 
is the universality of 
$R(c_{x})$, which  was verified by the authors  for quite a wide range of 
shaking intensities (temperatures) and  number of particles (densities) 
of the system. They present this as a robust property and, in fact,
it has been used as a guide to develop a model for vibrated granular systems
\cite{ByT02}. Nevertheless, recent experiments by Blair and Kudrolli
\cite{ByK01} suggest that the scaled horizontal VDF of a vibrated 
granular system depends on the number of particles, and only when this   
number is small enough it can be well fitted by $R(c_{x})$. 
It can be argued that the reason for this apparent contradiction is  that 
the experimental setup was different in both cases
and, in fact, inhomogeneities were stronger in the experiment by Blair and
Kudrolli than in the one by Rouyer and Menon. Whether these inhomogeneities
are enough to destroy the universal behavior found by Rouyer and Menon
is not clear at the moment.

In this paper, we will study the distribution function of the horizontal 
component of the velocity  of a vibrated granular fluid in the presence
of gravity by using Molecular Dynamics simulations. Computer simulations
are a very useful tool to study the velocity distribution of a 
system, as they provide  detailed information about positions and
velocities of all particles of the system at any moment. The construction 
of the corresponding distributions is straightforward, and averages
over different times and initial conditions provide the statistical
accuracy needed to compute the tails of the distributions. The goal
of this study is to clarify the behavior of the velocity
distribution of a fluidized low density granular system in the
presence of gravity, and, in particular, to investigate the existence of
a universal scaling behavior, independently of the parameters
of the system. 

\section{Molecular Dynamics Simulations}
\label{sec2}

We consider a two dimensional system of $N$ inelastic hard disks of 
diameter $\sigma$ and mass $m$, whose collisions are  characterized by
a constant coefficient of normal restitution $\alpha$. There is a gravity
field $\bf{g}=-g \bf{u}_{y}$, acting in the negative direction of the $Y$ 
axis, and a vibrating wall located at the bottom of the system ($y=0$). 
The system size in the $X$ direction is $S$, and periodic boundary conditions
are employed in that direction. Besides, and unless otherwise indicated,
the system is opened at the top. The vibrating wall is elastic, and moves
in a sawtooth manner, i.e., all particles colliding with the
wall find it moving upwards with velocity $v_{w}$. Moreover, the
amplitude of the wall vibration will be considered so small
as compared with the mean free path of the particles,  that
the position of the wall can be taken accurately as  fixed at $y=0$. Although 
this is clearly an ideal approximation to the movement of a real 
vibrating wall, we will be studying the VDF in regions not too close 
to the wall, so the details of its movement are expected to be unimportant.

	The described system  exhibits a steady state with gradients 
only in the direction of gravity which has been well  
characterized in the  low density limit\cite{BRyM01} 
by using a hydrodynamic description. This steady state  depends
both on $\alpha$ and $N_{y}=N/S$, which is proportional to the number
of monolayers in the direction of gravity at rest. 
It must be pointed out that this state is not always stable,
and inhomogeneities appear in the transversal direction if the
system is wide enough, both with and without gravity 
\cite{SyK01,BRMyG02,LMyS02}. 
In this work, we are interested  in the VDF of the transversally homogeneous 
state, so the parameters will be chosen to ensure that inhomogeneities
do not show up in the $X$-direction.

	The evolution of the system  was followed by
using standard event driven Molecular Dynamics simulations \cite{AyT87},
for several values of the parameters. In particular, the coefficient
of normal restitution was always in the range $\alpha \geq 0.9$, which
contains the value $\alpha=0.92$ characteristic of steel balls used
in most of the mentioned experiments. The system width $S$ was given
a value $S=50\sigma$, which was small enough as to ensure homogeneity in
the direction perpendicular to gravity for the values of $\alpha$
considered. The gravity field was set to $g=1$, while $m$ defines the unit
of mass and $\sigma$ the unit of length. The velocity of the 
vibrating  wall was chosen in each case  such that the number density, $n$, 
was always bellow a small enough value, $n\leq 0.15$, even in the denser 
regions of the system. Finally, for every value of $\alpha$, several different 
values of the number of particles were considered, in order to
study its possible influence on the VDF. Let us notice that, as the system
width $S$ is fixed, changing the number of particles is equivalent to
changing the number of monolayers $N_{y}$, which is the relevant
parameter.

	The simulations started from an arbitrary
initial situation that was left to evolve until the system reached
a steady state.  Typical density and temperature 
profiles in the steady state
are shown in Fig. \ref{fig1}. Although the details of the profiles
depend on the particular values of the parameters of the system,
in all the cases reported here the density goes 
through a maximum at a certain height, 
while the temperature reaches a minimum, increasing from there on.
It is important to notice that the temperature minimum does not coincide
with the density maximum, and that the condition of small density
gradient in a given region of the system does not imply small temperature 
gradient in it at all.
In order to avoid the effect of inhomogeneities in the VDF, we have 
considered narrow layers perpendicular to the direction
of the gradients at different positions throughout the system. 
The height of the cells $\Delta y_{c}$ 
was chosen so that the variation of density and temperature within them were 
negligible. For instance, in the case displayed in Fig. \ref{fig1}, the value
$\Delta y_{c}=5$ was used. Cells at different positions  will 
be considered to get  information about the velocity distribution 
throughout  the whole system.
The distribution function for $v_{x}$, $P_{x}(v_{x};y)$  in each cell
is constructed  by sampling the velocities of particles inside it 
at different times,  once the system was in the steady state. Besides,
also different trajectories starting from different initial conditions
were considered in the average process, in order to improve the statistics.

	As the system is inhomogeneous in the $Y$ direction,
$P_{x}(v_{x};y)$  is expected to depend on height, $y$. 
Previous studies suggest that this dependence occurs only 
through the second moment of the distribution, 
but this is a result that has not been proved up to now for
vibrated granular systems in presence of gravity. 
We  then  define a scaled velocity by
\begin{equation}   
\label{eq2:1}
c_{x}=\frac{v_{x}}{\langle v_{x}^{2}\rangle^{1/2}} \, ,
\end{equation}
and a scaled velocity distribution,
\begin{equation}
\phi_{x}(c_{x};y)=\langle v_{x}^{2}\rangle^{1/2}
P_{x}(v_{x};y) \, .
\label{eq2:2}
\end{equation}
If all the dependence of the distribution on height occurs through
the second moment (i.e. the temperature), $\phi_{x}$ must be
independent on height. This is one of the points to be checked in
the simulations.

\section{Scaled velocity distribution}
\label{sec3}

In Figs. \ref{fig2} and \ref{fig3} the scaled velocity distribution
$\phi_{x}$ for $\alpha=0.95$ and two different values 
of the number of particles,
$N=300$ ($N_{y}=6$) and $N=600$ ($N_{y}=12$), is plotted. The different 
symbols correspond to different heights in the system, as indicated
in the figures. The solid line is the gaussian distribution, while the 
dotted-dashed line is the stretched exponential $R(c_{x})$ defined in 
Eq.\ (\ref{eq1:1}). The layers whose distribution is displayed in the figures
are located before the density  maximum, at the maximum, and after it.
The position of the density maximum is $y\simeq 100$ 
for both number of particles.  
The reason to display both  $\phi_{x}$ and its logarithm is that the former
provides visual information about the distribution for thermal
velocities, while the latter is more convenient to identifying
the behavior of the tails of the distribution.

	The first point we want to investigate is whether the spatial
dependence of the velocity distribution can be scaled out with its (local)
second moment. If we consider the natural plot of $\phi_{x}$,
the answer to this question may seem affirmative, at least in very
good approximation. For given values of the parameters of the system,
the symbols corresponding to different heights fall onto the same curve.
Nevertheless, if we pay attention to the tails of the distribution,
the answer is not so clear. Closer inspection of the logarithmic
plots of  $\phi_{x}$ shows that the scaling is not
exact, and, in particular, the distribution corresponding to the
layer located at the density maximum has always the most populated tail
and, therefore, is the narrowest distribution. This implies that the
height dependence of the distribution has not been completely scaled out.
Besides, the differences in shape of  $\phi_{x}$ at different heights
cannot be a consequence of  averaging in space an inhomogeneous 
distribution, since the height of the layers is very small, as
indicated above. Still, it could happen that
the deviations from the scaling behavior were only relevant  in a narrow
region around the density maximum. To investigate whether this is the case, 
and also to provide a more quantitative measurement of the departure from the 
scaling, we have computed the scaled fourth moment of the distribution, 
$\gamma_{4}$, defined as
\begin{equation}
\gamma_{4}=\frac{\langle v_{x}^{4} \rangle}{\langle v_{x}^{2} \rangle^{2}}
\, .
\end{equation}
 If the velocity distribution depended on height only through its
second moment, $\gamma_{4}$ would be constant through the
system. If, in addition, the velocity distribution were a gaussian, 
$\gamma_{4}=3$, while if it were given by the stretched
exponential $R(c_{x})$,   $\gamma_{4}\simeq 3.756$. 

In Figs. \ref{fig4} and \ref{fig5}
$\gamma_{4}$ has been plotted as a function of height for $\alpha=0.95$
and $\alpha=0.9$, respectively. In each case,  several values of the number 
of particles in the system have been considered. 
The range of $y$ plotted corresponds to the relevant region in the system,
i.e.,  before the density decays to a very low value.	
In all cases, $\gamma_{4}$ exhibits a clear dependence on height.
For given $\alpha$, the failure of the
scaling is more evident the larger the number of particles in the
system. In fact, it is the number of particles and not the density the 
parameter governing the behavior of  the VDF,  as far as the system 
is fluidized. This has been checked by changing $v_{w}$ for fixed
$\alpha$ and $N$, which implies modifying  the density of the system,
since it is opened at the top. The curves obtained for $\gamma_{4}$ 
were identical if the height was properly scaled with the vibrating 
velocity being used.  The quantity $\gamma_{4}$ always follows the same
trend: it takes a value very close to the gaussian one in the
vicinity of the vibrating wall, then passes trough a maximum, and decays to
a more or less constant value that depends on both 
$\alpha$ and $N$. Comparison of the behavior of $\gamma_{4}$ 
with the density and temperature profiles 
in each case shows that the position of the maximum of 
$\gamma_{4}$ is close to the density maximum, while the  
region with a roughly constant value of the scaled fourth moment 
occurs beyond the temperature minimum. It must be noticed as well that, 
for the largest number of monolayers considered,
$\gamma_{4}$ seems to increase for large heights. Although it is true that the
statistical noise there is large, increasing the statistics does not change 
the observed behavior.

The first conclusion emerging  from the simulations is that the
dependence on height of the transversal VDF of a vibrated granular
gas in the presence of gravity cannot be scaled out 
with its width, i.e., with the local transversal granular temperature
of the system. This might appear as rather surprising,
especially taking into account what happens in other
states of a granular gas. For instance, in the homogeneous cooling
state,  the VDF depends on time,  but this dependence
can be completely scaled out with the temperature,
and the scaled fourth velocity moment reaches a plateau after
a transient period \cite{BRyC96}. On the other hand, it is  true that,
for the cases reported here,
the scaling can be taken as an accurate  approximation in the
thermal velocity region, at least for a not too large number of particles 
in the system.  Besides, the more inelastic the system,
the smaller the number of monolayers  for which deviations from scaling 
begin to be significant.

	Figures \ref{fig2}-\ref{fig5} also indicate that, 
even if we neglect the height dependence,  the shape
of the distribution still depends both on the inelasticity and the
number of monolayers. For instance, in the case of Fig. $\ref{fig2}$,
the VDF is very well approximated by a gaussian distribution
for velocities up to $|c_{x}|\sim 4$. Some deviations from the
gaussian appear for large velocities, but they are not very significant.
As the number of particles is increased, keeping $\alpha$ fixed,
the deviations from the gaussian shape become large, and the 
distribution gets narrower. This is the case of Fig. \ref{fig3}, where
$\phi_{x}$ lies between the gaussian distribution and $R(c_{x})$
for thermal velocities, but the tail of the distribution
is better approximated by the latter.  If the number of particles
is further increased, the distribution becomes narrower, 
in fact, narrower than  $R(c_{x})$. Again, as it happened with the scaling,
the smaller $\alpha$, the smaller the number of particles for which deviations
from the gaussian distribution appear, as it follows from Figs. \ref{fig4} and 
\ref{fig5}.

Some authors have argued that, in a vibrated granular system, 
collisions with the wall tend to randomize the VDF, rendering it wide, 
while collisions between  particles act as the ordering factor, because 
particle  velocities are more parallel  after collisions than 
before \cite{KyH00}. To check if this argument is consistent
with our simulation results,  we have computed $R_{pp/pw}$, the ratio 
between  the number of particle-particle  and particle-wall collisions. 
We found that, as expected, it is an increasing function of the number of 
monolayers in the system. For instance, for  $\alpha=0.95$ we
found $R_{pp/pw}\sim 41.1$ for $N=300$,  $R_{pp/pw}\sim 67.5$ for $N=400$,
and  $R_{pp/pw}\sim 154.2$ for $N=600$. Since, as indicated above, the
width of the distribution decreases as the number of particles increases,
we conclude that the larger $R_{pp/pw}$,
the narrower the VDF. Moreover, the more inelastic the system,
the more parallel the velocities after collisions, and the  
``ordering'' mechanism is expected to be more efficient. This explains why,
although for   $\alpha=0.9$ and $N=410$ it is $R_{pp/pw}\sim 103.3$,
the distribution  is in this case narrower than  the narrowest 
reported here for $\alpha=0.95$, which corresponds to larger values
of the ratio  $R_{pp/pw}$.

Finally,  the fact that the accuracy of the fit 
of the VDF to both $R(c_{x})$ and  the gaussian distribution 
depends in most of the cases  on the velocity interval considered,
implies that $\phi_{x}$  cannot be 
fitted by a  stretched exponential
function $\phi_{x}\sim\exp\{-C |c_{x}|^{\beta}\}$, with a single
value of the exponent $\beta$, in the whole velocity range.

All the above discussion  suggests, in our opinion, that the role played 
by the gaussian
distribution and $R(c_{x})$ in the context of vibrated granular gases
is very different. The gaussian distribution is a limiting behavior
for small enough number of particles and not too inelastic systems.
As the number of particles in the system increases, the scaled velocity 
distribution becomes narrower. In this range of parameters,
$R(c_{x})$ is  just a good fit  valid in some cases, 
but it does not represent a universal behavior, not even a limiting
behavior of the VDF for  some well-defined conditions. 

\section{Symmetric boundary conditions}
\label{sec4}

Given that the conclusions reached up to now in this work 
are so different from those emerging from the  
experiments reported  by Rouyer and Menon \cite{RyM00}, it seems 
worth  to simulate the same kind of boundary conditions they use, 
in order to discard that the discrepancies are due to peculiarities of
the boundaries. Therefore, instead of considering just one vibrating 
wall at $y=0$, we have added
another vibrating wall at $y=L=50\sigma$. The choice of the maximum
height, $L$, of the system, was made as to obtain very similar system
dimensions as those in Ref.  \cite{RyM00}. The two vibrating
walls are identical, and the vibrating velocity is chosen to generate 
symmetric around the center of the system stationary 
density and temperature profiles.
The total number of particles was varied in the range $300\leq N\leq 500$. 
In each case, we started from a homogeneous state, that was left to
evolve until a stationary situation was reached.  
Then, the transversal VDF was computed in a layer located in the the center
of the system, namely  between $0.4L$ and $0.6L$. 
Care was taken that the system remained in a fluidized state in that 
central region. Note that, in this way and as it was done in 
Ref. \cite{RyM00}, the VDF is measured now in a quite wide region of the 
system, and not in a narrow layer as we did in the previous study of 
the open system. Even if the gradients in that central region 
are expected to be small, they will be non vanishing, and the
nonuniformity might influence the shape of the VDF.

In Fig. \ref{fig6} we have plotted the scaled distribution 
$\phi_{x}$ for $\alpha=0.92$, which is roughly the value of
the coefficient of restitution for the steel balls used in \cite{RyM00},
 and three different values of the
number of particles. As it was the case for  the open system,
the scaled VDF is seen to depend on the number of particles,  
becoming narrower as $N$ increases. This can be interpreted again
in terms of the role of the particle-particle  versus
particles-wall number of collisions, as also in this case the ratio $R_{pp/pw}$
is an increasing function of the number of monolayers. 
More specifically, we found from the simulation data that 
$R_{pp/pw}\sim 42$ for $N=300$, $R_{pp/pw}\sim 68$ for $N=400$, 
and $R_{pp/pw}\sim 103$ for $N=500$. 

	The relevant role of the stretched exponential $R(c_{x})$ is
also questioned by Fig. \ref{fig6}. In all the cases displayed, 
the distribution for thermal velocities lies between the gaussian and 
$R(c_{x})$. The
tail of the distribution gets closer to  $R(c_{x})$ the larger $N$,
and in fact only in the case $N=500$  is  well fitted
by the stretched exponential. Therefore, again the distribution cannot
be fitted by a single stretched exponential in the whole velocity
range. This is more clearly realized in  Fig. \ref{fig7}, where  
$\ln(-\ln (\phi_{x}/\phi_{x}(0))$ 
has been plotted as a function of $\ln c_{x}$.
Let us notice that, if the distribution were some general  stretched
exponential, $\phi_{x}\sim\exp\{-C |c_{x}|^{\beta}\}$, this plot would give 
a straight line whose slope would provide  the exponent $\beta$. The symbols
in the figure are from the simulations. Also plotted are two 
straight lines, one corresponding to $\beta=2$ (gaussian behavior)
and the other one to $\beta=1.5$ ($R(c_{x})$) as a guide to the eye.
A transition from gaussian behavior for thermal velocities
to a distribution with a smaller exponent $\beta$ 
for larger velocities is observed in all cases. 

	The results discussed up to now correspond to the VDF in the central
region of the system. Nevertheless, we expect  the transversal
scaled VDF to depend also on $y$  in this case. In Fig. \ref{fig8}, 
$\gamma_{4}$ has been plotted
as a function of height for the same values of the parameters
as in Figs. \ref{fig6} and \ref{fig7}. Again, we find that the dependence
on height and the deviations from the gaussian are larger the larger
the number of monolayers. As expected,  $\gamma_{4}$
is symmetric around the center of mass. The exhibited double peak 
clearly indicated  that the 
maximum of $\gamma_{4}$ does not happen either at
the density maximum or at the temperature minimum, 
since they are both located at the center of the system.

\section{Conclusions}
\label{sec5}

The transversal velocity distribution of a vibrated granular gas in the 
presence of gravity has been studied by using Molecular Dynamics
simulations. It has been found that the shape of the distribution depends
both on the inelasticity of the system and the number of monolayers
in the direction of gravity. Besides, it depends on height,
and  this dependence cannot be fully scaled out with
its width, i.e., with the local transversal granular temperature of the
system. This contrasts with some previous reported results  
for granular gases under similar conditions. On the other hand,
it is true that the scaling provides a quite accurate 
approximation valid for not too inelastic systems and for not too large
number of monolayers at rest.

For a  given value of the coefficient of restitution $\alpha$, the transversal
velocity distribution becomes narrower as the number of monolayers in
the system is increased. This can be understood by taking into account that
increasing the number of monolayers  increases the ratio of number of 
particle-particle collisions against particle-wall collisions.
In a granular system, the particle-particle collisions tend to suppress
``noise'' in the velocity distribution, as the velocities after the
collision are more parallel than before it. Then, it is sensible
to expect that, the larger the relative number of particle-particle collisions,
the narrower the velocity distribution. On the other hand,
the more inelastic the system the more effective each collision
in narrowing the velocity distribution.

	The role played by the gaussian distribution and the
stretched exponential $R(c_{x})$ has also been clarified. While
the gaussian is a limiting behavior shown in the quasi-elastic and
small number of monolayers limit, $R(c_{x})$ appears  just as a
good fit for the velocity distribution for some range of values
of the parameters characterizing the system. 

\acknowledgments

We acknowledge support from the Ministerio de Ciencia y Tecnolog\'{\i}a
(Spain) through Grant No. BFM2002-00303 (partially financed by
FEDER founds).

\begin{figure}
\caption{Steady density (solid line) and temperature (dotted-dashed
line) profiles for a system with $\alpha=0.95$ and $N_{y}=6$. The temperature
has been scaled with some arbitrary value, while density and height are 
measured in the dimensionless units defined in the main text.}
\label{fig1}
\end{figure}

\begin{figure}
\caption{Scaled velocity distribution at different heights 
for $\alpha=0.95$ and $N_{y}=6$ (the
same values of Fig. \ref{fig1}). The symbols are from the simulations,
the continuous line is for a  gaussian distribution, and the dotted-dashed
line is for the stretched exponential $R(c_{x})$.}
\label{fig2}
\end{figure}

\begin{figure}
\caption{The same as Fig. \ref{fig2} but for $\alpha=0.95$ and $N_{y}=12$.
The layer at $y=100$ corresponds to the density maximum, which is
$n=0.15$ in this case.}
\label{fig3}
\end{figure}

\begin{figure}
\caption{Scaled fourth moment as a function of height for $\alpha=0.95$
and three different values of the number of monolayers. The height 
has been divided by $y_{M}$, which is defined in each case as
a height such that the density has decayed to a very low value.}
\label{fig4}
\end{figure}

\begin{figure}
\caption{The same as Fig. \ref{fig4} but for $\alpha=0.9$.}
\label{fig5}
\end{figure}

\begin{figure}
\caption{Scaled velocity distribution at the central region of the system
for symmetric boundary conditions, $\alpha=0.92$ and three values of the
number of particles. The symbols are from the simulations, the
continuous line is the gaussian distribution, and the dotted-dashed line 
$R(c_{x})$.}
\label{fig6}
\end{figure}

\begin{figure}
\caption{The same as in Fig. \ref{fig6}, but using a double logarithmic
plot. The solid line has the slope that corresponds to a gaussian 
distribution, while the dotted-dashed line has the slope associated
to $R(c_{x})$. The positions of this two lines has been chosen quite 
arbitrarily  in order to fit the maximum range of the simulation
results.}
\label{fig7}
\end{figure}

\begin{figure}
\caption{Scaled fourth moment as a function of height in the symmetric
case for the same values of the parameters as in Figs. \ref{fig6}
and \ref{fig7}.}
\label{fig8}
\end{figure}

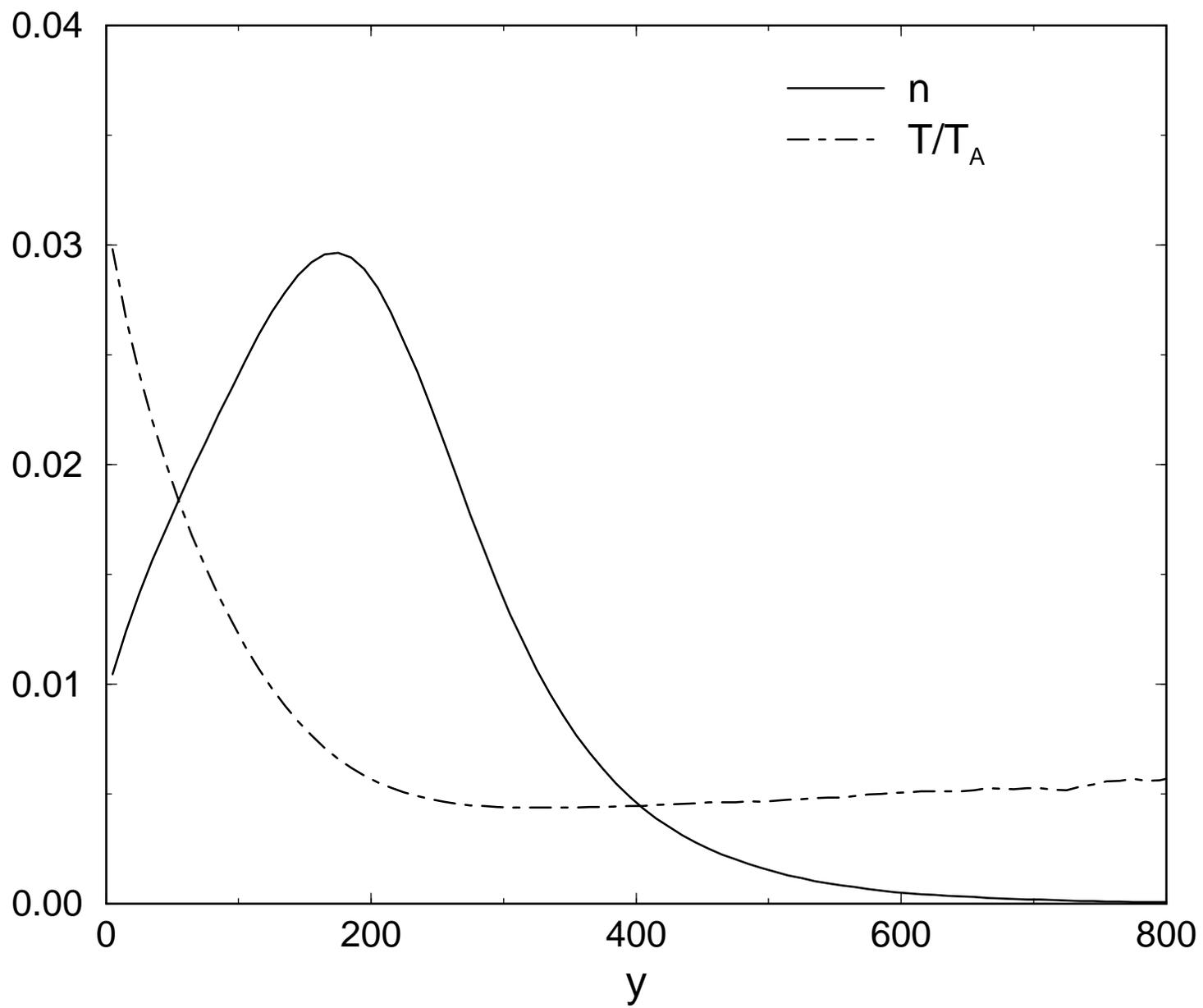
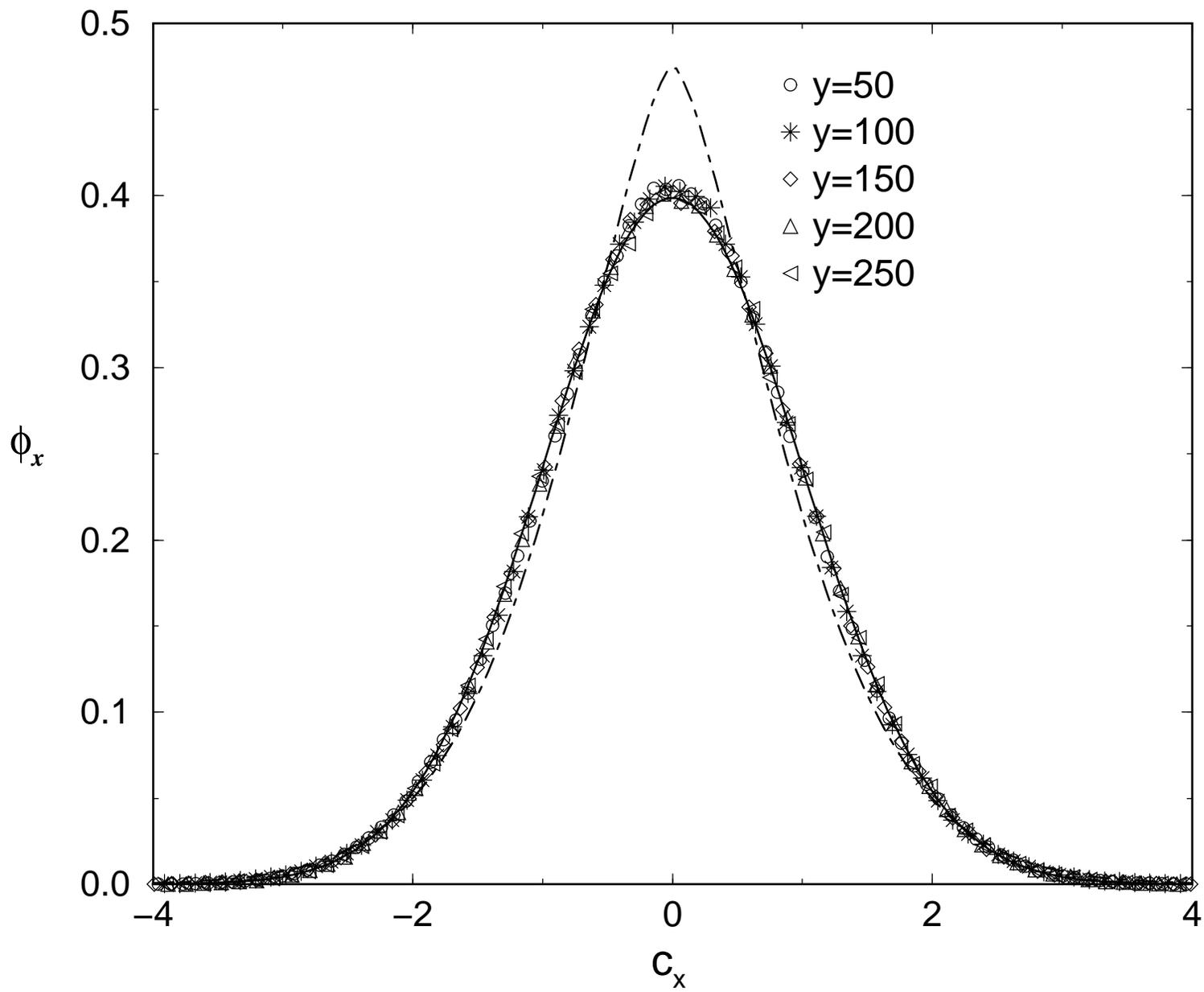
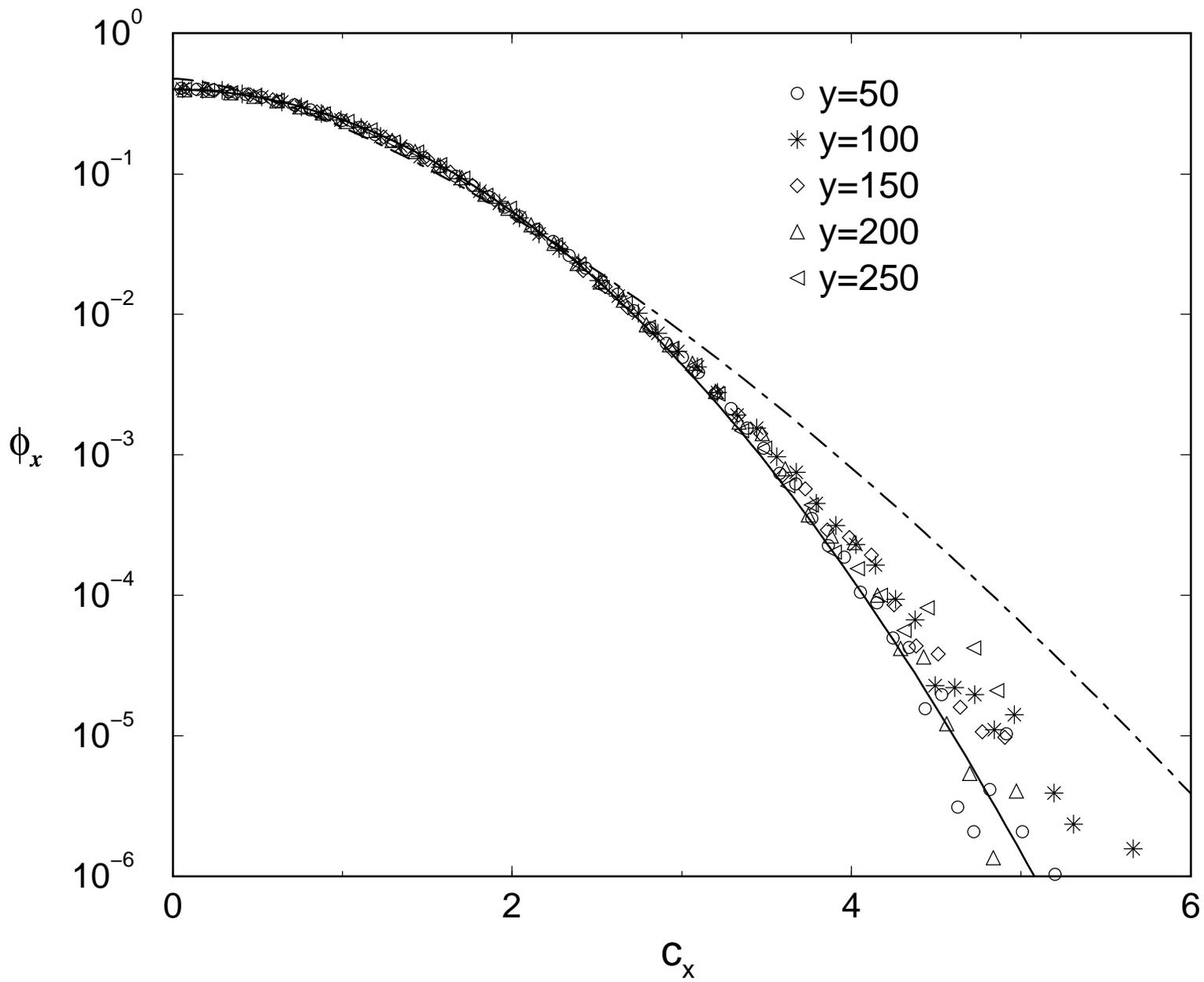
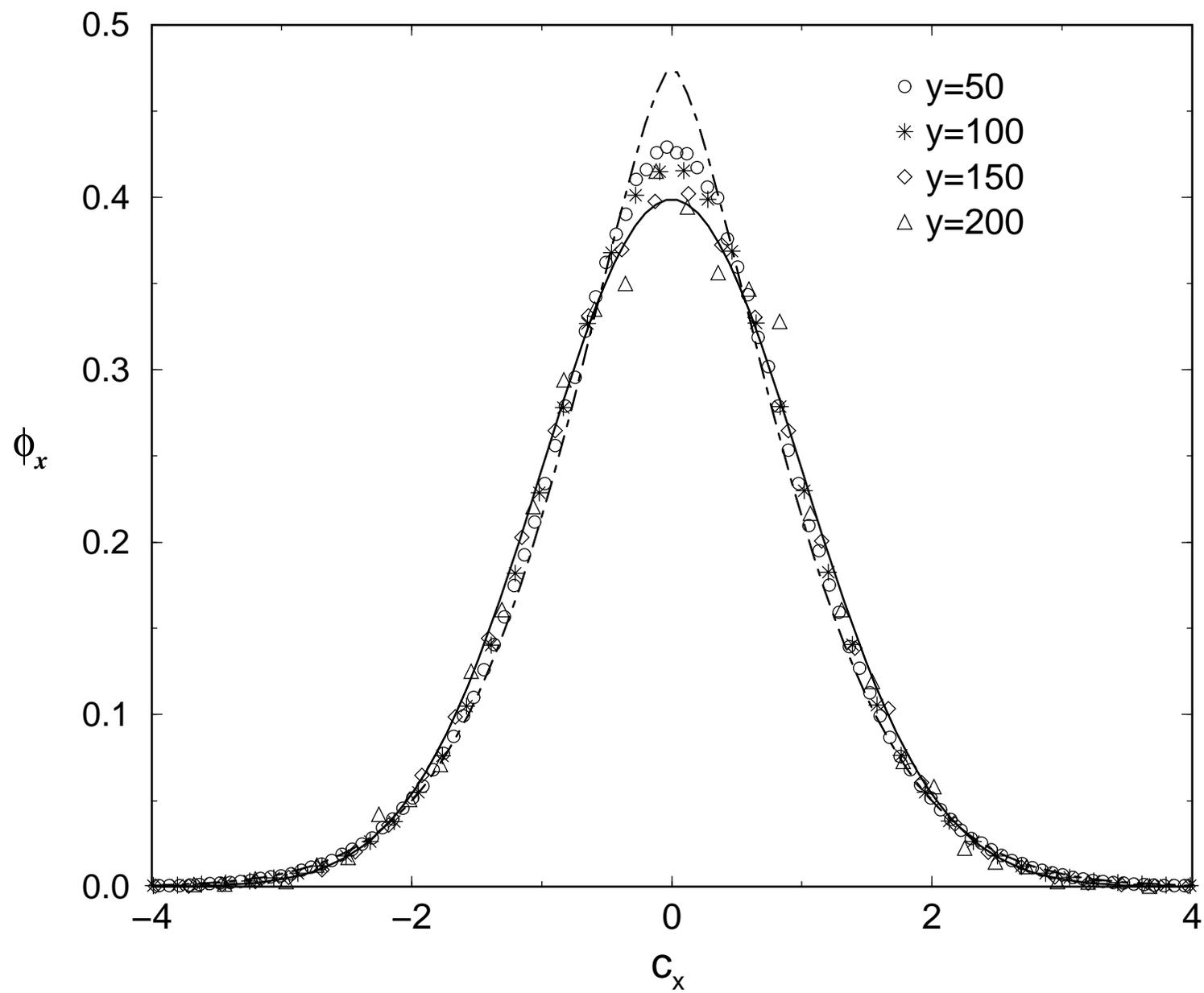
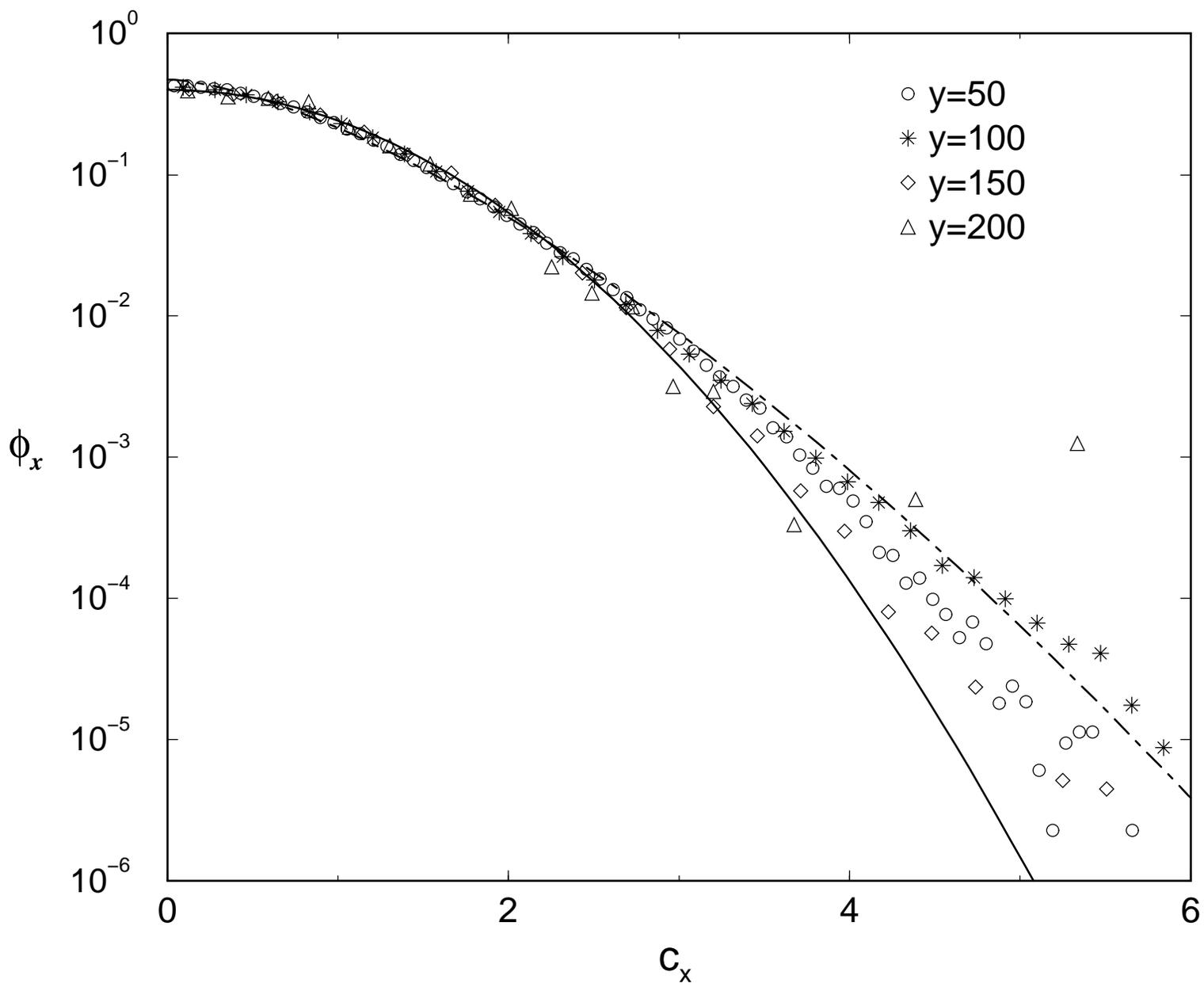
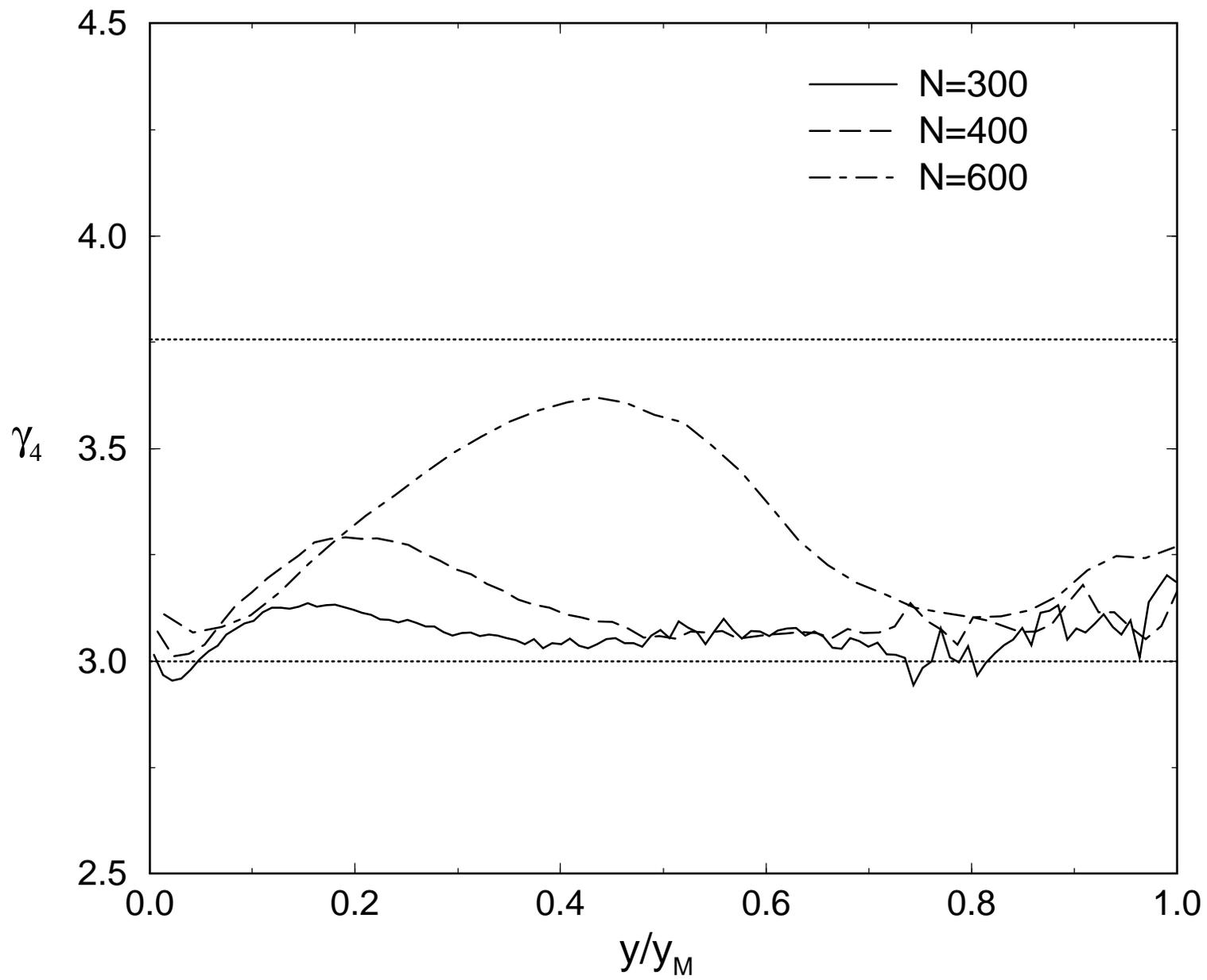
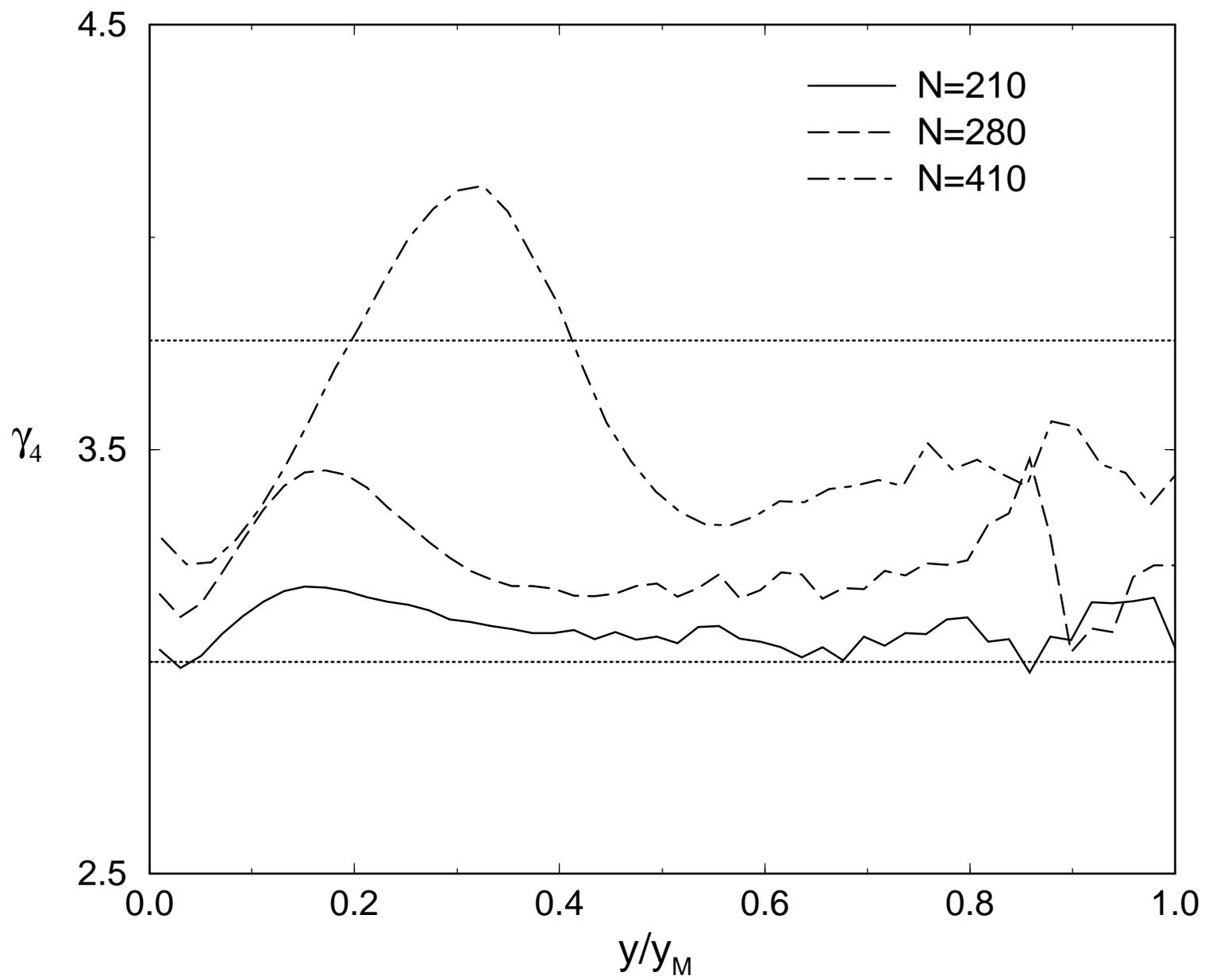
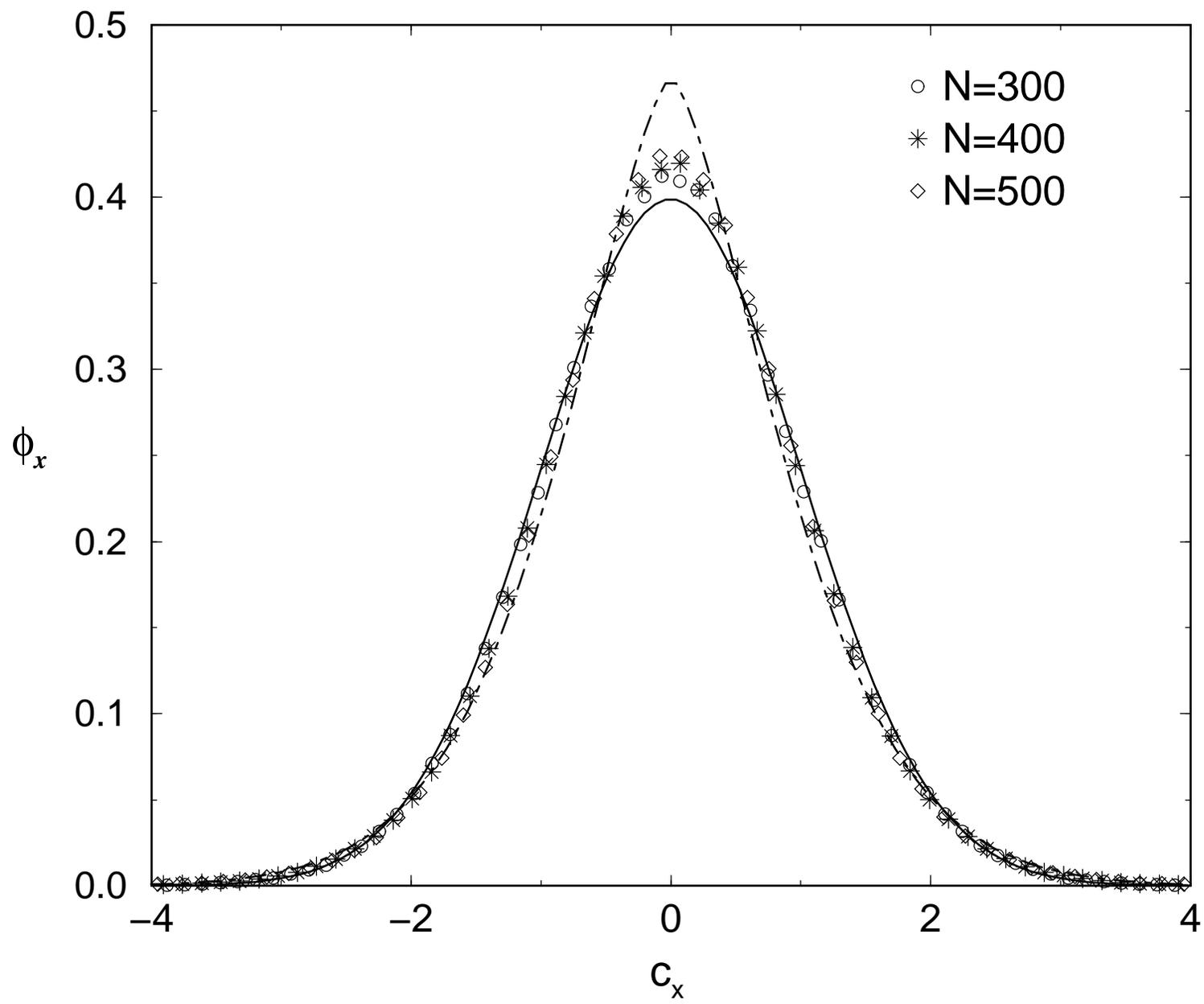
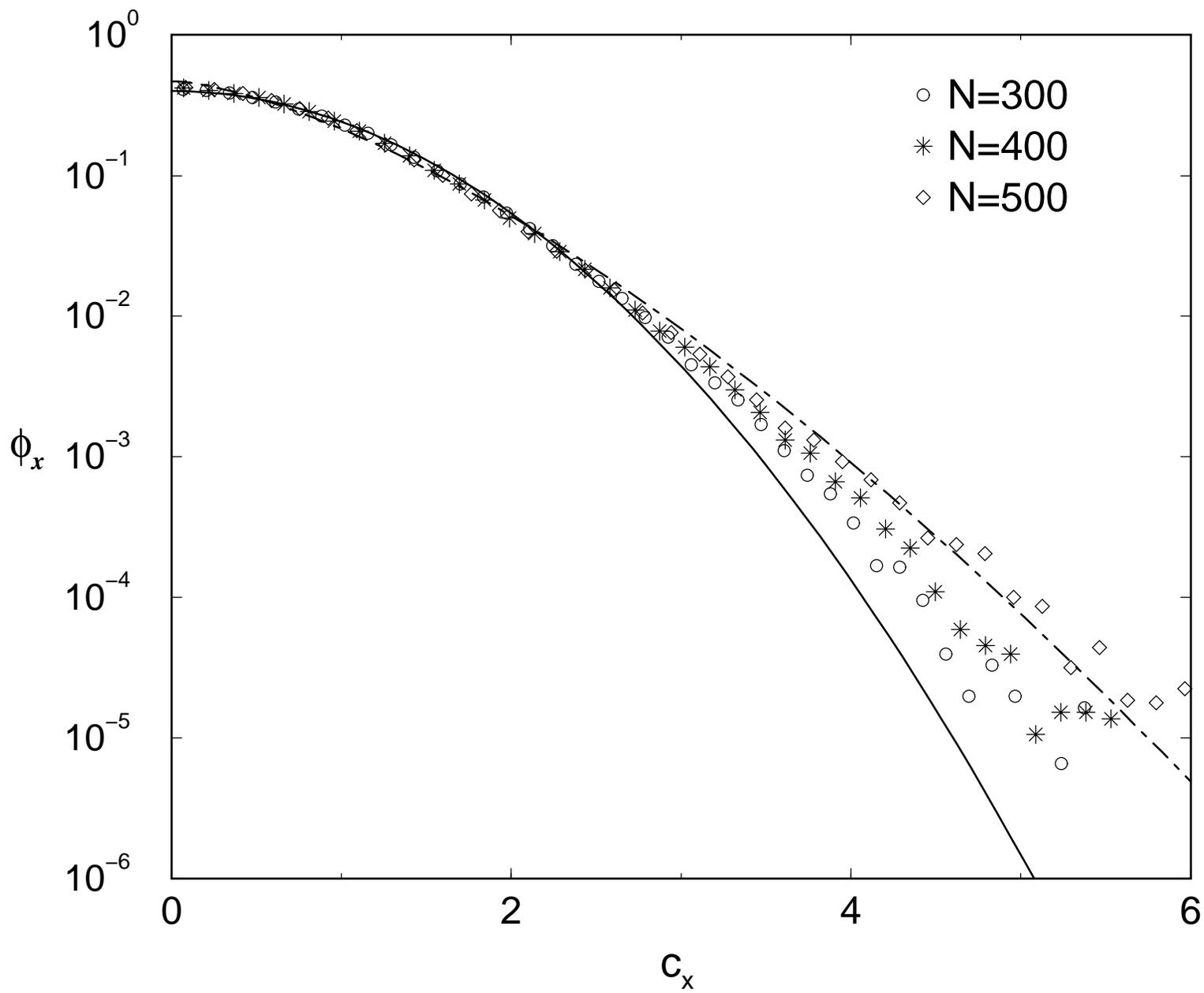
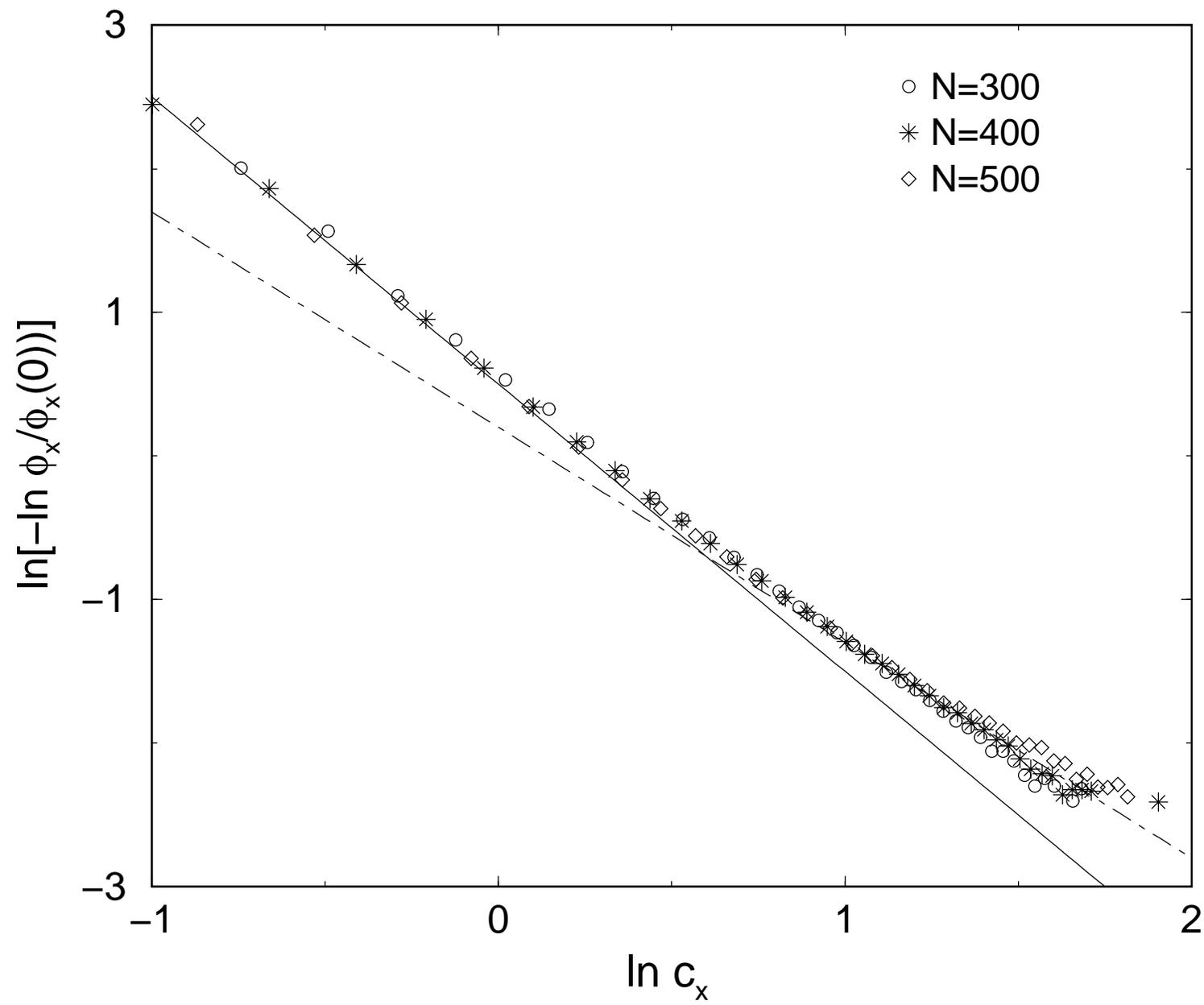
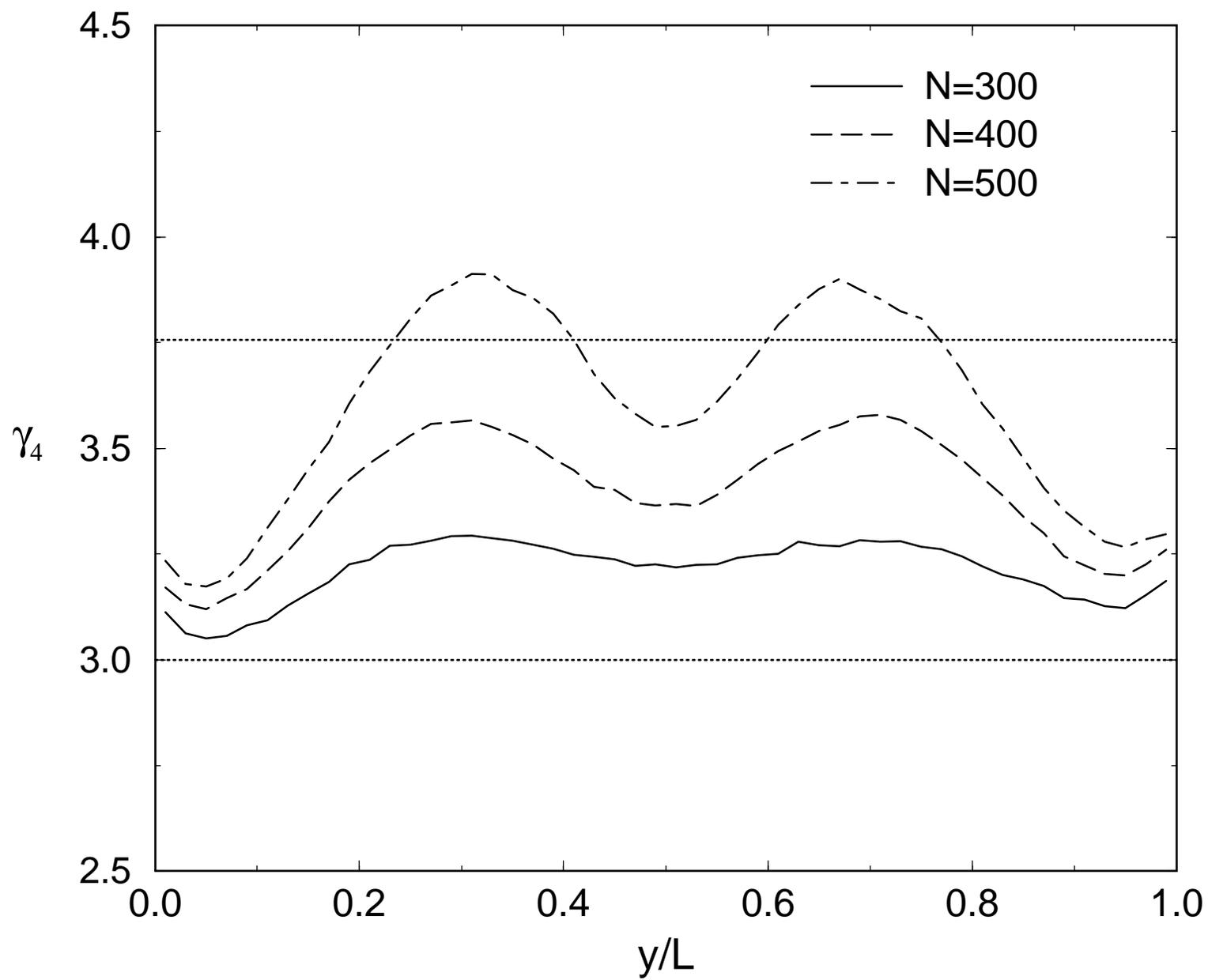
\end{document}